\documentclass[preprint,10pt]{elsarticle}

\makeatletter
\def\ps@pprintTitle{%
 \let\@oddhead\@empty
 \let\@evenhead\@empty
 \def\@oddfoot{\centerline{\thepage}}%
 \let\@evenfoot\@oddfoot}
\makeatother

\usepackage{graphicx}
\usepackage{subcaption}
\usepackage[utf8]{inputenc}
\usepackage{lscape}
\usepackage{booktabs}
\usepackage[normalem]{ulem}
\useunder{\uline}{\ul}{}
\usepackage{amsmath}

\usepackage[a4paper,margin=1.3in]{geometry}
\usepackage{amssymb}
\usepackage{lineno}

\journal{a journal}

\begin{document}

\begin{frontmatter}

%% Title, authors and addresses

\title{Automatic Detection of Coronavirus Disease (COVID-19) in X-ray and CT Images: A Machine Learning Based Approach}

%% use the tnoteref command within \title for footnotes;
%% use the tnotetext command for the associated footnote;
%% use the fnref command within \author or \address for footnotes;
%% use the fntext command for the associated footnote;
%% use the corref command within \author for corresponding author footnotes;
%% use the cortext command for the associated footnote;
%% use the ead command for the email address,
%% and the form \ead[url] for the home page:

%****************************************************************
%****************************************************************
%****************************************************************

% \author[First]{Sara Hosseinzadeh Kassani\corref{cor1}\fnref{label2}}
\author[First]{Sara Hosseinzadeh Kassani\corref{cor1}}
\ead{sara.kassani@usask.ca}
% \ead[url]{www.cea-ifac.es}

\author[Second]{Peyman Hosseinzadeh Kassasni}
\ead{peymanhk@stanford.edu}
% \ead[url]{www2.cea-ifac.es}

\author[Third]{Michal J. Wesolowski}
% \ead{autor3@cea-ifac.es}

\author[First]{Kevin A. Schneider}
% \ead{autor3@cea-ifac.es}

\author[First]{Ralph Deters}
% \ead{autor3@cea-ifac.es}

% \fntext[label2]{Nota al pie para el autor 1}
\cortext[cor1]{Corresponding author.}

\address[First]{Department of Computer Science, University of Saskatchewan, Saskatchewan, Canada}
\address[Second]{Department of Neurology and Neurological, University of Stanford, California, United States}
\address[Third]{Department of Medical Imaging, University of Saskatchewan, Saskatchewan, Canada}
%****************************************************************
%****************************************************************
%****************************************************************

\begin{abstract}
%% Text of abstract
The newly identified Coronavirus pneumonia, subsequently termed COVID-19, is highly transmittable and pathogenic with no clinically approved antiviral drug or vaccine available for treatment. The most common symptoms of COVID-19 are dry cough, sore throat, and fever. Symptoms can progress to a severe form of pneumonia with critical complications, including septic shock, pulmonary edema, acute respiratory distress syndrome and multi-organ failure. While medical imaging is not currently recommended in Canada for primary diagnosis of COVID-19, computer-aided diagnosis systems could assist in the early detection of COVID-19 abnormalities and help to monitor the progression of the disease, potentially reduce mortality rates. In this study, we compare popular deep learning-based feature extraction frameworks for automatic COVID-19 classification. To obtain the most accurate feature, which is an essential component of learning, MobileNet, DenseNet, Xception, ResNet, InceptionV3, InceptionResNetV2, VGGNet, NASNet were chosen amongst a pool of deep convolutional neural networks. The extracted features were then fed into several machine learning classifiers to classify subjects as either a case of COVID-19 or a control. This approach avoided task-specific data pre-processing methods to support a better generalization ability for unseen data. The performance of the proposed method was validated on a publicly available COVID-19 dataset of chest X-ray and CT images. The DenseNet121 feature extractor with Bagging tree classifier achieved the best performance with 99\% classification accuracy. The second-best learner was a hybrid of the a ResNet50 feature extractor trained by LightGBM with an accuracy of 98\%.
\end{abstract}

\begin{keyword}
Coronavirus Disease \sep Lung Opacity \sep Computer-Aided Diagnosis\sep Deep Learning \sep Feature Extraction \sep Transfer Learning
%% keywords here, in the form: keyword \sep keyword

%% MSC codes here, in the form: \MSC code \sep code
%% or \MSC[2008] code \sep code (2000 is the default)

\end{keyword}

\end{frontmatter}

%%
%% Start line numbering here if you want
%%
% \linenumbers

%% main text
\section{Introduction}
\label{S:Introduction}

A series of pneumonia cases of unknown etiology occurred in December 2019, in Wuhan, Hubei province, China. On December 31, 2019, 27 unexplained cases of pneumonia were identified and found to be associated with so called “wet markets” which sell fresh meat and seafood from a variety of animals including bats and pangolins. The pneumonia was found to be caused by a virus identified as "severe acute respiratory syndrome coronavirus 2” (SARS-CoV-2),  with the associated disease subsequently termed coronavirus disease 2019 (COVID-19) by the World Health Organization (WHO)~\cite{SHEREEN202091}~\cite{LIPPI2020145}. Genomic analysis showed that COVID-19 is phylogenetically related to SARS-like bat viruses. Hence, bats could be the possible source of the viral replication~\cite{SHEREEN202091}. Pangolins have also been identified as a potential intermediate host of COVID-19~\cite{ZHANG20201346}. 
This newly identified virus is highly transmittable and pathogenically different from SARS-CoV, MERS-CoV, avian influenza, influenza, and other common respiratory viruses. Concerning the outbreak of COVID-19, on January 30, 2020, the WHO declared the outbreak of the novel Coronavirus disease as a Public Health Emergency of International Concern (PHEIC)~\cite{GUO2020}. The rapid worldwide spread of disease resulted in a global pandemic declaration on March 11, 2020.  Clinical symptoms of patients infected with COVID-19 are similar to other viral upper respiratory diseases such as Influenza, respiratory syncytial virus (RSV), and bacterial pneumonia. The most common presenting symptoms experienced by patients include dry cough, sore throat, fever, dyspnea, diarrhea, myalgia, shortness of breath and bilateral lung infiltrates, observable on clinical imaging such as chest X-ray. Other symptoms are headache, vomiting, pleurisy, sneezing, rhinorrhea, and nasal congestion. Patients with more severe COVID-19 have developed critical complications, including septic shock, pulmonary edema, cardiac injury, acute kidney injury, Acute Respiratory Distress Syndrome (ARDS) and even Multi-Organ Failure (MOF)~\cite{GUO2020}~\cite{CHAVEZ2020}.
At present, there is no clinically approved antiviral drug or vaccine available to treat COVID-19. The reproduction number (R0), defined as the expected number of susceptible cases directly generated by one infectious case of COVID-19 infection, is estimated to 3.77~\cite{ROTHAN2020102433}~\cite{LIU2020}. Despite global efforts of travel restrictions and quarantine, while the epidemic continues to decline in China, the incidence of novel COVID-19 continues to rise globally, with over 1.6 million confirmed cases and over 100,000 deaths worldwide, at the time of this writing~\cite{CoronavirusStat}. As of April 2020, substantial new incidence of COVID-19 cases have been reported in 211 countries with significant confirmed cases in South Korea, Italy, Iran, Japan, Germany, and France~\cite{SHIM2020339}. The early spread of new COVID-19 cases was associated with recent travel to China; however, community spread is now common globally. The greatest number of new cases occur through close contact human-to-human transmission (approximately 6 feet) by respiratory droplets~\cite{CHAVEZ2020}. Contamination also can occur through infected surfaces with subsequent contact with the eyes, nose, or mouth. 

\begin{figure}[ht]
\centering\includegraphics[width=0.6\linewidth]{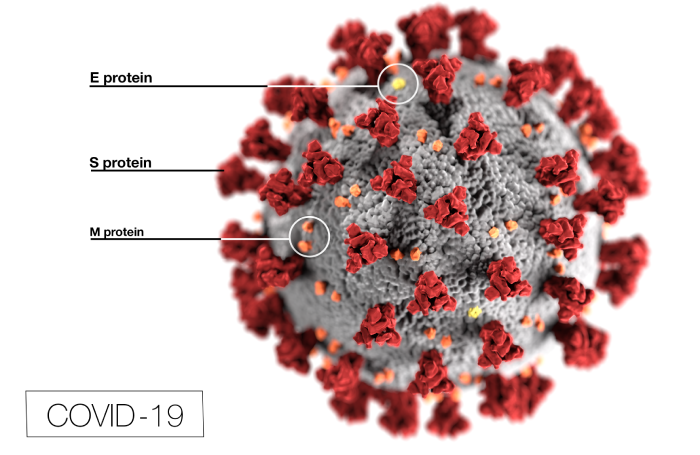}
\caption{The illustration of COVID-19, created at the Centers for Disease Control and Prevention (CDC)~\cite{cdcWebLink1}. The protein particles E, S, and M are located on the outer surface of the virus particle.}
    \label{fig:covid1}
\end{figure}

\begin{figure}[ht]
\centering\includegraphics[width=0.6\linewidth]{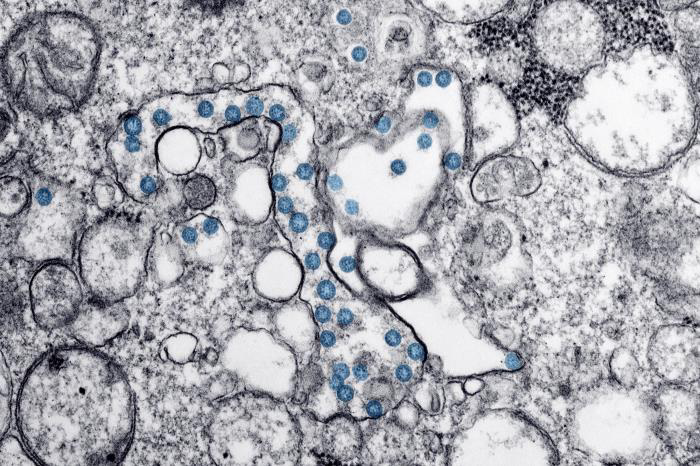}
\caption{Transmission electron microscopic image of a case of COVID-19. The spherical viral particles, colorized blue, contain cross-sections through the viral genome, seen as black dots~\cite{cdcWebLink2}.}
    \label{fig:covid2}
\end{figure}

The genetic characteristics of the Coronavirus should be well understood to fight against this virus. Coronavirus is a single-stranded RNA virus consisting of approximately 27–32 kb with particle size ranged from 65-125nm in diameter~\cite{SHEREEN202091}. An illustration of COVID-19 is shown in Figure~\ref{fig:covid1}. A transmission electron microscopic image of a case of COVID-19 is also demonstrated in Figure~\ref{fig:covid2}.

In light of this, it is evident that early detection of COVID-19 is necessary to interrupt the spread of COVID-19 and prevent transmission by early isolation of patients, trace and quarantine of close contacts. In patients with COVID-19, accurate monitoring of the disease progression is a critical component of disease management. While not recommended for primary diagnosis of COVID-19 in Canada, medical imaging modalities such as chest X-ray and Computed Tomography (CT) play an important role in confirming diagnosis of positive COVID-19 pneumonia as well as monitoring the progression of the disease. These types of images show an extent of irregular ground-glass opacities that progress rapidly after COVID-19 symptom onset.  These abnormalities peaked during days 6-11 of the illness. The second most predominant pattern of lung opacity abnormalities peak during days 12-17 of the illness~\cite{wang2020temporal}. Computer-Aided Diagnosis (CAD) systems that incorporate X-ray and CT image processing techniques and deep learning algorithms could assist physicians as diagnostic aides for COVID-19 and help provide a better understanding of the progression the disease. 

\subsection{Related Research}
\label{S:RelatedResearch }

Hemdan et al.~\cite{hemdan2020covidx} developed a deep learning framework, COVIDX-Net, to diagnose COVID-19 in X-Ray Images. A comparative study of different deep learning architectures including VGG19, DenseNet201, ResNetV2, InceptionV3, InceptionResNetV2, Xception and MobileNetV2 is provided by authors. The public dataset of X-ray images was provided by Dr. Joseph Cohen~\cite{cohen2020covid} and Dr. Adrian Rosebrock~\cite{AdrianRosebrock}. The provided dataset included 50 X-ray images, divided into two classes as 25 normal cases and 25 positive COVID-19 images. Hemdan’s results demonstrated VGG19 and DenseNet201 models achieved the best performance scores among counterparts with 90.00\% accuracy.

Barstugan et al.~\cite{barstugan2020coronavirus} proposed a machine learning approach for COVID-19 classification from CT images. Patches with different sizes 16$\times$16, 32$\times$32, 48$\times$48, 64$\times$64 were extracted from 150 CT images. Different hand-crafted features such as Grey Level Co-occurrence Matrix (GLCM), Local Directional Pattern (LDP), Grey Level Run Length Matrix (GLRLM), Grey-Level Size Zone Matrix (GLSZM), and Discrete Wavelet Transform (DWT) algorithms were employed. The extracted features were fed into a Support Vector Machine (SVM)~\cite{cristianini2000introduction} classifier on 2-fold, 5-fold and 10-fold cross-validations. The best accuracy of 98.77\% was obtained by GLSZM feature extractor with 10-fold cross-validation.

Wang and Wong~\cite{wang2020covid} designed a tailored deep learning-based framework, COVID-Net, developed for COVID-19 detection from chest X-ray images. The COVID-Net architecture was constructed of combination of 1$ \times $1 convolutions, depth-wise convolution and the residual modules to enable design deeper architecture and avoid the gradient vanishing problem. The provided dataset consisted of s a combination of COVID chest X-ray dataset provided by Dr. Joseph Cohen~\cite{cohen2020covid}, and Kaggle chest X-ray images dataset~\cite{kagglechestxray} for a multi-class classification of normal, bacterial infection, viral infection (non-COVID) and COVID-19 infection. Obtained accuracy of this study was 83.5\%.

In a study conducted by Maghdid et al.~\cite{maghdid2020diagnosing}, a deep learning-based method and transfer learning strategy were used for automatic diagnosis of COVID-19 pneumonia. The proposed architecture is a combination of a simple convolutional neural network (CNN) architecture (one convolutional layer with 16 filters followed by batch normalization, rectified linear unit (ReLU), two fully-connected layers) and a modified AlexNet~\cite{krizhevsky2012imagenet} architecture with the feasibility of transfer learning. The proposed modified architecture achieved an accuracy of 94.00\%.

Ghoshal and Tucker~\cite{ghoshal2020estimating} investigated the diagnostic uncertainty and interpretability of deep learning-based methods for COVID-19 detection in X-ray images. Dropweights based Bayesian Convolutional Neural Networks (BCNN) were used to estimate uncertainty in deep learning solutions and provide a level of confidence of a computer-based diagnosis for a trusted clinician setting. To measure the relationship between accuracy and uncertainty, 70 posterioranterior (PA) lung X-ray images of COVID-19 positive patients from the public dataset provided by Dr. Joseph Cohen~\cite{cohen2020covid} were selected and balanced by Kaggle’s Chest X-Ray Images dataset~\cite{kagglechestxray}. To prepare the dataset, all images were resized to 512$\times$512 pixels. A transfer learning strategy and real-time data augmentation strategies were employed to overcome the limited size of the dataset. The proposed Bayesian inference approach obtained the detection accuracy of 92.86\% on X-ray images using VGG16 deep learning model.

Hall et al.~\cite{hall2020finding} used a VGG16 architecture and transfer learning strategy with 10-fold cross-validation trained on the dataset from Dr. Joseph Cohen~\cite{cohen2020covid}. All images were rescaled to 224$\times$224 pixels and a data augmentation strategy was employed to increase the size of dataset. The proposed approach achieved an overall accuracy 96.1\% and overall Area Under Curve (AUC) of 99.70\% on the provided dataset.

Farooq and Hafeez~\cite{farooq2020covid} proposed a fine-tuned and pre-trained ResNet-50 architecture, COVID-ResNet, for COVID-19 pneumonia screening. To improve the generalization of the training model, different data augmentation methods including vertical flip, random rotation (with angle of 15 degree), along with the model regularization were used. The proposed method achieved the accuracy of 96.23\% on a multi-class classification of normal, bacterial infection, viral infection (non-COVID-19) and COVID-19 infection dataset.

\subsection{Motivation and contributions}
\label{S:contributions}

The main motivation of this study is to present a generic feature extraction method using convolutional neural networks that does not require handcrafted or very complex features from input data while being easily applied to different modalities such as X-ray and CT images. Another primary goal is to reduce the generalization error while achieving a more accurate diagnosis. The contributions are summarized as follows:
\begin{itemize}
\item Deep convolutional feature representation~\cite{boureau2010theoretical,rakhlin2018deep,guo2016deep} is used to extract highly representative features using state-of-the-art deep CNN descriptors. The employed approach is able to discriminate between COVID-19 and healthy subjects from chest X-ray and CT images and hence produce higher accuracy in comparison to other works presented in the literature. To the best of our knowledge, this research is the first comprehensive study of the application of machine learning (ML) algorithms (15 deep CNN visual feature extractor and 6 ML classifier) for automatic diagnoses of COVID-19 from X-ray and CT images.

\item To overcome the issue of over-fitting in deep learning due to the limited number of training images, a transfer-learning strategy is adopted as the training of very deep CNN models from scratch requires a large number of training data.

\item No data augmentation or extensive pre-processing methods are applied to the dataset in order to increase the generalization ability and also reduce bias toward the model performance.

\item The proposed approach reduces the detection time dramatically while achieving satisfactory accuracy, which is a superior advantage for developing real or near real-time inferences on clinical applications. 

\item With extensive experiments, we show that the combination of a deep CNN with Bagging trees classifier achieves very good classification performance applied on COVID-19 data despite the limited number of image samples.

\item Finally, we developed an end to end web-based detection system to simulate a virtual clinical pipeline and facilitate the screening of suspicious cases.
\end{itemize}

The rest of this paper is organized as follows. The proposed methodology for automatically classifying COVID-19 and healthy cases is explained in Section~\ref{S:ProposedMethodology}. The dataset description, experimental settings and performance metrics are given in Section~\ref{S:Experiments}. A brief discussion and results analysis are provided in Section~\ref{S:Discussion}, and finally, the conclusion is presented in Section~\ref{S:Conclusion}.

\section{Proposed Methodology}
\label{S:ProposedMethodology}
Few studies have been published on the application of deep CNN feature descriptors to X-ray and CT images. Each of the CNN architectures is constructed by different modules and convolution layers that aid in extracting fundamental and prominent features from a given input image. Briefly, in the first step, we collect available public chest X-ray and CT images. In the next step, we pre-processed the provided dataset using standard image normalization techniques to improve the quality of visual information of the input data. Once input images are prepared, we fed them into the feature extraction phase with the state-of-the-art CNN descriptors to extract deep features from each input image. For the training phase, the generated features are then fed into machine learning classifiers such as Decision Tree (DT)~\cite{quinlan1986induction}, Random Forest (RF)~\cite{Breiman2001}, XGBoost~\cite{Chen2016}, AdaBoost~\cite{freund1995desicion}, Bagging classifier~\cite{breiman1996bagging} and LightGBM~\cite{ke2017lightgbm}. Finally, the performance of the proposed approach is evaluated on test images.

\subsection{Feature extraction using transfer learning}
\label{S:FeatureExtraction}
The concept of transfer learning has been introduced for solving deep learning problems arising from insufficiently labeled data, or when the CNN model is too deep and complex. Aiming to tackle these challenges, studies in a variety computer vision tasks demonstrated the advantages of transfer learning strategies from an auxiliary domain in improving the detection rate and performance of a classifier~\cite{kassani2019breast}~\cite{khan2019novel}~\cite{mehra2018breast}. In a transfer learning strategy, we transfer the weights already learned on a cross-domain dataset into the current deep learning task instead of training a model from scratch. With the transfer learning strategy, the deep CNN can obtain general features from the source dataset that cannot be learned due to the limited size of the dataset in the current task. Transfer learning strategies have various advantages, such as avoiding the overfitting issue when the number of training samples is limited, reducing the computational resources, and also speeding up the convergence of the network~\cite{lu2019pathological}~\cite{biopsybinary}.

\subsection{CNN Descriptor}
\label{S:CNNDescriptor}
\begin{figure}[ht!]
\centering\includegraphics[width=0.99\linewidth]{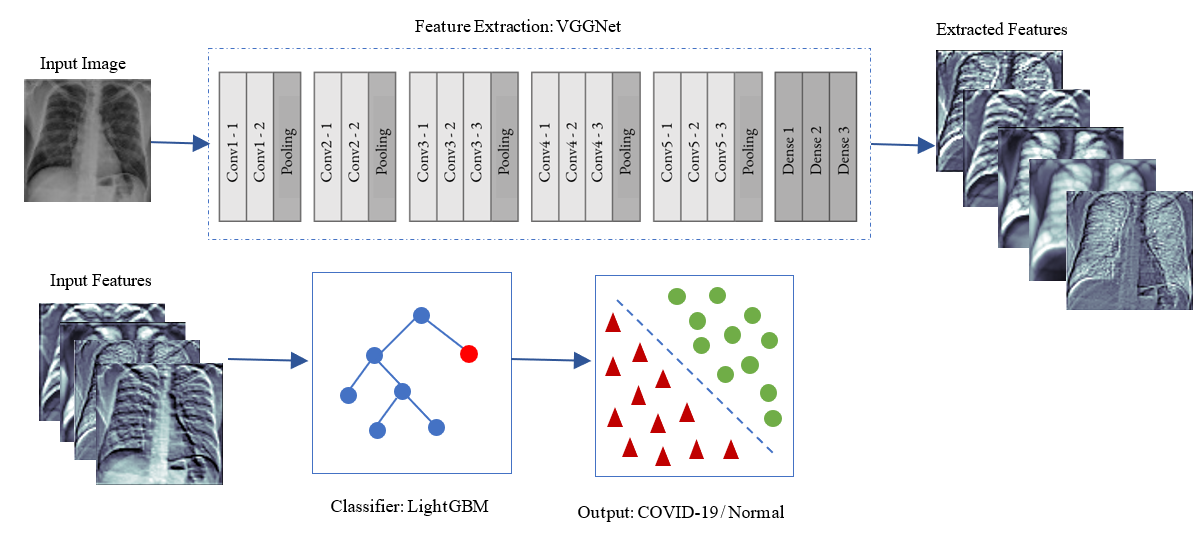}
\caption{General framework of the proposed method with VGGNet as feature extractor.}\label{covidArchitecture}
\end{figure}

Effective feature extraction is one of the most important steps toward learning rich and informative representations from raw input data to provide accurate and robust results. The small or imbalanced size of the training samples poses a significant challenge for the training of a deep CNN where data dimensionality is much larger than the number of samples leading to over-fitting. Although various strategies, e.g. data augmentation~\cite{liu2020automatic}, transfer learning~\cite{liu2019novel} and fine-tuning~\cite{sridar2019decision}, may reduce the problem of insufficient or imbalance training data, the detection rate of the CNN model may degrade due to the over-fitting issue. Since the overall performance obtained by a fine-tuning method in the initial experiments for this study was not significant, we employed a different approach inspired by~\cite{boureau2010theoretical}~\cite{rakhlin2018deep}~\cite{guo2016deep} known as deep convolutional feature representation. In this method, we used pre-trained well-established CNN models as a visual feature extractor to encode the input images into a feature vector of sparse descriptors of low dimensionality. Then the computed encoded feature vectors produced by CNN architectures are fed into different classifiers, i.e. machine learning algorithms, to yield the final prediction. This lower dimension vector significantly reduces the risk of over-fitting and also the training time. Different robust CNN architectures such as MobileNet, DenseNet, Xception, InceptionV3, InceptionResNetV2, ResNet, VGGNet, NASNet are selected  for feature extraction with the possibility of transfer learning advantage for limited datasets and also their satisfying performances in different computer vision tasks~\cite{zhang2019automated,dourado2019deep,ccinar2020detection,cogan2019mapgi}. Figure~\ref{covidArchitecture}. illustrates the visual features extracted by VGGNet architecture from an X-ray image of a COVID-19 positive patient.

\section{Experiments}
\label{S:Experiments}
\subsection{Dataset description}
\label{S:Dataset}
\begin{figure}[ht!]
	\centering
	
	\begin{subfigure}{0.22\textwidth}
		\includegraphics[width=\linewidth]{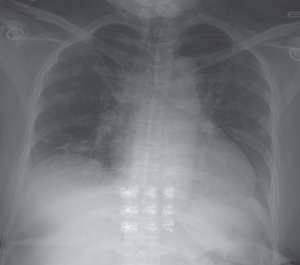}
		\caption{}
% 		\label{fig:original}
	\end{subfigure}\hfil
	\begin{subfigure}{0.22\textwidth}
		\includegraphics[width=\linewidth]{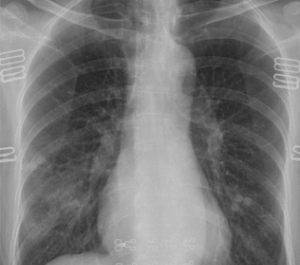}
		\caption{}
% 		\label{fig:VF}
	\end{subfigure}\hfil
	\begin{subfigure}{0.22\textwidth}
		\includegraphics[width=\linewidth]{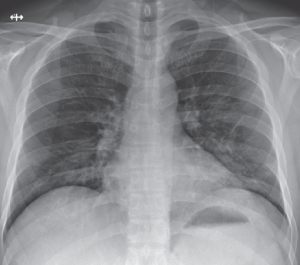}
		\caption{}
% 		\label{fig:HF}
	\end{subfigure}\hfil	
	\begin{subfigure}{0.22\textwidth}
		\includegraphics[width=\linewidth]{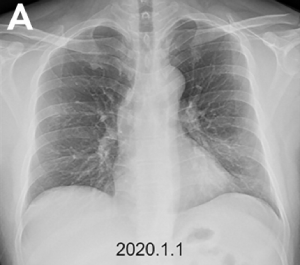}
		\caption{}
% 		\label{fig:RF}

	\end{subfigure}\hfil

	\caption{Chest X-ray images of four confirmed COVID-19 pneumonia. (a) 52-year old female, presenting diffuse infiltrates in the bilateral lower lungs. (b) 59-year old female, demonstrating right infahilar airspace opacities. (c) 35-year old male, presenting stable streaky opacities in the lung bases, indicating likely atypical pneumonia; the opacities have steadily increased in density over time. (d) 42-year old male, presenting opacities in the left lower and right upper lobes on day 7 after the onset of symptoms.}
	\label{fig:dataExamples}
\end{figure}

In order to evaluate the performance of our feature extracting and classifying approach, we used the public dataset of X-ray images provided by Dr. Joseph Cohen available from a GitHub repository~\cite{cohen2020covid}. We used the available 117 chest X-ray images and 20 CT images (137 images in total) of COVID-19 positive cases. We also included 117 images of healthy cases of X-ray images from Kaggle Chest X-Ray Images (Pneumonia) dataset available at~\cite{kagglechestxray} and 20 images of healthy cases of CT images from Kaggle RSNA Pneumonia Detection dataset available at~\cite{kaggleRSNA} to balance the dataset with both positive and normal cases. Figure~\ref{fig:dataExamples} shows examples of confirmed COVID-19 images extracted from the provided dataset. The X-ray images of confirmed COVID-19 infection demonstrate different shapes of “pure ground glass” also known as hazy lung opacity with irregular linear opacity depending the disease progress~\cite{wang2020temporal}. 

\subsection{Data pre-processing}
\label{S:preprocessing}

The images within the dataset were collected from multiple imaging clinics with different equipment and image acquisition parameters; therefore, considerable variations exist in images’ intensity. The proposed method in this study avoids extensive pre-processing steps to improve the generalization ability of the CNN architecture. This helps to make the model more robust to noise, artifacts and variations in input images during feature extraction phase. Hence, we only employed two standard pre-processing steps in training deep learning models to optimize the training process.
\begin{itemize}
\item Resizing: The images in this dataset vary in resolution and dimension, ranging from 365$\times$465 to 1125$\times$859 pixels; therefore, we re-scaled all images of the original size to the size of 600$\times$450 pixels to obtain a consistent dimension for all input images. The input images were also separately resized to 331$\times$331 pixels and 224$\times$224 pixels as required for NASNetLarge and NASNetMobile architectures, respectively.

\item Image normalization: For image normalization, first, we re-scaled the intensity values of the pixels using ImageNet mean subtraction as a pre-processing step. The ImageNet mean is a pre-computed constant derived from the ImageNet database~\cite{krizhevsky2012imagenet}. Another essential pre-process step is intensity normalization. To accomplish this, we normalized the intensity values of all images from [0, 255] to the standard normal distribution by min-max normalization to the intensity range of [0, 1], which is computed as: 
\end{itemize}

 \begin{equation}\label{normalization}
  x{_{norm}} = \frac{x-x_{min}}{x_{max} - x_{min}}
\end{equation}

where $x$ is the pixel intensity. $x_{min}$ and $x_{max}$ are minimum and maximum intensity values of the input image in equation~\ref{normalization}. This operation helps to speed up the convergence of the model by removing the bias from the features and achieve a uniform distribution across the dataset. 

\subsection{Evaluation criteria}
\label{S:EvaluationCriteria}

To measure the prediction performance of the methods in this study, we utilized common evaluation metrics such as recall, precision, accuracy and f1-score. According to equations (2–5) True positive (TP) is the number of instances that correctly predicted; false negative (FN) is the number of instances that incorrectly predicted. True negative (TN) is the number of negative instances that predicted correctly, while false positive (FP) is the number of negative instances incorrectly predicted. Given TP, TN, FP and FN, all evaluation metrics were calculated as follows: 

Recall or sensitivity is the measure of COVID-19 cases that are correctly classified. Recall is critical, especially in the medical field and is given by: 

\begin{equation}
   Recall = \frac{TP}{TP+FN}
\end{equation}

Precision or positive predictive value is defined as the percentage of correctly classified labels in truly positive patients and is given as: 

\begin{equation}
   Precision = \frac{TP}{TP+FP}
\end{equation}

Accuracy shows the number of correctly classified cases divided by the total number of test images, and is defined as:

 \begin{equation}
    Accuracy = \frac{TP+TN}{TP+TN+FP+FN} 
 \end{equation}

F1-score, also known as F-measure, is defined as the weighted average of precision and recall that combines both the precision and recall together. F-measure is expressed as:

 \begin{equation}
   F1-score=2 \times \frac{Recall \times Precision}{Recall + Precision}
\end{equation}

\section{Discussion}
\label{S:Discussion}
Diagnostic imaging modalities, such as chest radiography and CT are playing an important role in confirming the primary diagnosis from the Polymerase Chain Reaction (PCR) test for COVID-19. Medical imaging is also playing a critical in monitoring the progression of the disease and patient care. Extracting features from radiology modalities is an essential step in training machine learning models since the model performance directly depends on the quality of extracted features. Motivated by the success of deep learning models in computer vision, the focus of this research is to provide an extensive comprehensive study on the classification of COVID-19 pneumonia in chest X-ray and CT imaging using features extracted by the state-of-the-art deep CNN architectures and trained on machine learning algorithms. The 10-fold cross-validation technique was adopted to evaluate the average generalization performance of the classifiers in each experiment. For all CNNs, the network weights were initialized from the weights trained on ImageNet. The Windows based computer system used for this work had an Intel(R) Core(TM) i7-8700K 3.7 GHz processors with 32 GB RAM. The training and testing process of the proposed architecture for this experiment was implemented in Python using Keras package with Tensorflow backend as the deep learning framework backend and run on Nvidia GeForce GTX 1080 Ti GPU with 11GB RAM. 

\begin{table}[ht!]
\caption{Comparison of classification performance ($\mu$ $\pm$ $\sigma$) of different machine learning models measured by accuracy. The bold value indicates the best result; underlined value represents the second-best result of the respective category.}
\label{tab:accuracy}
\centering
\resizebox{\textwidth}{!}{%
\begin{tabular}{@{}lllllll@{}}

\toprule
\textbf{}                  & \textbf{Decision Tree} & \textbf{Random Forest} & \textbf{XGBoost} & \textbf{AdaBoost} & \textbf{Bagging} & \textbf{LightGBM} \\ \midrule
\textbf{MobileNet}         & 83.00 $\pm$ 0.26           & 93.00 $\pm$ 0.23           & 95.00 $\pm$ 0.16     & 80.00 $\pm$ 0.17      & 96.00 $\pm$ -0.11    & 82.00 $\pm$ 0.28      \\ \midrule
\textbf{DesnseNet121}      & 92.00 $\pm$ 0.15           & 90.00 $\pm$ 0.21           & 94.00 $\pm$ 0.16     & 92.00 $\pm$ 0.19      & \textbf{99.00 $\pm$ 0.07}     & 96.00 $\pm$ 0.11      \\ \midrule
\textbf{DenseNet201}       & 84.00 $\pm$ 0.26           & 90.00 $\pm$ 0.24           & 90.00 $\pm$ 0.18     & 87.00 $\pm$ 0.25      & 96.00 $\pm$ 0.11     & 87.00 $\pm$ 0.17      \\ \midrule
\textbf{Xception}          & 95.00 $\pm$ 0.17           & 90.00 $\pm$ 0.19           & 96.00 $\pm$ 0.11     & 93.00 $\pm$ 0.20      & 96.00 $\pm$ 0.11     & 96.00 $\pm$ 0.11      \\ \midrule
\textbf{InceptionV3}       & 82.00 $\pm$ 0.22           & 84.00 $\pm$ 0.29           & 88.00 $\pm$ 0.15     & 80.00 $\pm$ 0.12      & 95.00 $\pm$ 0.12     & 84.00 $\pm$ 0.16      \\ \midrule
\textbf{InceptionResNetV2} & 84.00 $\pm$ 0.31           & 93.00 $\pm$ 0.16           & 93.00 $\pm$ 0.19     & 87.00 $\pm$ 0.33      & 94.00 $\pm$ 0.12     & 88.00 $\pm$ 0.21      \\ \midrule
\textbf{ResNet50}          & 89.00 $\pm$ 0.17           & 90.00 $\pm$ 0.15           & 93.00 $\pm$ 0.16     & 94.00 $\pm$ 0.12      & 93.00 $\pm$ 0.16     & {\ul 98.00 $\pm$ 0.09}      \\ \midrule
\textbf{ResNet152}         & 93.00 $\pm$ 0.12           & 92.00 $\pm$ 0.16           & 93.00 $\pm$ 0.16     & 94.00 $\pm$ 0.17      & 91.00 $\pm$ 0.22     & 93.00 $\pm$ 0.20      \\ \midrule
\textbf{VGG16}             & 90.00 $\pm$ 0.19           & 91.00 $\pm$ 0.19           & 88.00 $\pm$ 0.19     & 90.00 $\pm$ 0.19      & 90.00 $\pm$ 0.19     & 85.00 $\pm$ 0.19      \\ \midrule
\textbf{VGG19}             & 90.00 $\pm$ 0.19           & 87.00 $\pm$ 0.21           & 88.00 $\pm$ 0.19     & 90.00 $\pm$ 0.19      & 90.00 $\pm$ 0.19     & 85.00 $\pm$ 0.25      \\ \midrule
\textbf{NASNetLarge}       & 82.00 $\pm$ 0.23           & 88.00 $\pm$ 0.19           & 89.00 $\pm$ 0.17     & 81.00 $\pm$ 0.23      & 93.00 $\pm$ 0.19     & 82.00 $\pm$ 0.26      \\ \midrule
\textbf{NASNetMobile}      & 87.00 $\pm$ 0.17           & 88.00 $\pm$ 0.22           & 94.00 $\pm$ 0.19     & 87.00 $\pm$ 0.17      & 93.00 $\pm$ 0.19     & 89.00 $\pm$ 0.17      \\ \midrule
\textbf{ResNet50V2}        & 87.00 $\pm$ 0.12           & 96.00 $\pm$ 0.11           & 92.00 $\pm$ 0.19     & 90.00 $\pm$ 0.18      & 95.00 $\pm$ 0.12     & 88.00 $\pm$ 0.10      \\ \midrule
\textbf{ResNet101V2}       & 79.00 $\pm$ 0.32           & 89.00 $\pm$ 0.24           & 89.00 $\pm$ 0.28     & 76.00 $\pm$ 0.32      & 95.00 $\pm$ 0.12     & 78.00 $\pm$ 0.26      \\ \midrule
\textbf{ResNet152V2}       & 90.00 $\pm$ 0.27           & 86.00 $\pm$ 0.26           & 93.00 $\pm$ 0.16     & 89.00 $\pm$ 0.20      & 96.00 $\pm$ 0.11     & 88.00 $\pm$ 0.28      \\ \bottomrule
\end{tabular}%
}
\end{table}

\begin{table}[ht!]
\centering
\caption{Comparison of classification precision metric of different machine learning models. The bold value indicates the best result; underlined value represents the second-best result of the respective category.}
\label{tab:precision}
\resizebox{\textwidth}{!}{%
\begin{tabular}{@{}lllllll@{}}
\toprule
                           & \textbf{Decision Tree} & \textbf{Random Forest} & \textbf{XGBoost} & \textbf{AdaBoost} & \textbf{Bagging Classifier} & \textbf{LightGBM} \\ \midrule
\textbf{MobileNet}         & 89.00\%                & 88.00\%                & 93.00\%          & 85.00\%           & \textbf{99.00\%}            & 90.00\%           \\ \midrule
\textbf{DesnseNet121}      & 96.00\%                & 97.00\%                & {\ul 98.00\%}    & 95.00\%           & 96.00\%                     & 95.00\%           \\ \midrule
\textbf{DenseNet201}       & 94.00\%                & 94.00\%                & 95.00\%          & 94.00\%           & {\ul 98.00\%}               & 94.00\%           \\ \midrule
\textbf{Xception}          & 92.00\%                & 95.00\%                & 90.00\%          & 89.00\%           & {\ul 98.00\%}               & 93.00\%           \\ \midrule
\textbf{InceptionV3}       & 85.00\%                & 85.00\%                & 96.00\%          & 85.00\%           & \textbf{99.00\%}            & 82.00\%           \\ \midrule
\textbf{InceptionResNetV2} & 88.00\%                & 96.00\%                & 95.00\%          & 90.00\%           & 95.00\%                     & 93.00\%           \\ \midrule
\textbf{ResNet50}          & 95.00\%                & 89.00\%                & 94.00\%          & 96.00\%           & 95.00\%                     & 94.00\%           \\ \midrule
\textbf{ResNet152}         & 90.00\%                & 91.00\%                & 95.00\%          & 91.00\%           & 93.00\%                     & 89.00\%           \\ \midrule
\textbf{VGG16}             & 94.00\%                & 93.00\%                & 94.00\%          & 89.00\%           & 92.00\%                     & 89.00\%           \\ \midrule
\textbf{VGG19}             & 94.00\%                & 93.00\%                & 94.00\%          & 89.00\%           & 92.00\%                     & 89.00\%           \\ \midrule
\textbf{NASNetLarge}       & 89.00\%                & 91.00\%                & 94.00\%          & 90.00\%           & 95.00\%                     & 91.00\%           \\ \midrule
\textbf{NASNetMobile}      & 89.00\%                & 87.00\%                & 95.00\%          & 88.00\%           & 93.00\%                     & 88.00\%           \\ \midrule
\textbf{ResNet50V2}        & 92.00\%                & 89.00\%                & 94.00\%          & 88.00\%           & 96.00\%                     & 91.00\%           \\ \midrule
\textbf{ResNet101V2}       & 87.00\%                & 89.00\%                & 94.00\%          & 86.00\%           & 96.00\%                     & 78.00\%           \\ \midrule
\textbf{ResNet152V2}       & 91.00\%                & 94.00\%                & 96.00\%          & 91.00\%           & 97.00\%                     & 91.00\%           \\ \bottomrule
\end{tabular}%
}
\end{table}

\begin{table}[ht!]
\centering
\caption{Comparison of classification recall metric of different machine learning models. The bold value indicates the best result; underlined value represents the second-best result of the respective category.}
\label{tab:recall}
\resizebox{\textwidth}{!}{%
\begin{tabular}{@{}lllllll@{}}
\toprule
\textbf{}                  & \textbf{Decision Tree} & \textbf{Random Forest} & \textbf{XGBoost} & \textbf{AdaBoost} & \textbf{Bagging Classifier} & \textbf{LightGBM} \\ \midrule
\textbf{MobileNet}         & 89.00\%                & 88.00\%                & 93.00\%          & 84.00\%           & \textbf{99.00\%}            & 90.00\%           \\ \midrule
\textbf{DesnseNet121}      & 96.00\%                & 96.00\%                & {\ul 98.00\%}    & 95.00\%           & 96.00\%                     & 95.00\%           \\ \midrule
\textbf{DenseNet201}       & 94.00\%                & 94.00\%                & 95.00\%          & 94.00\%           & {\ul 98.00\%}               & 94.00\%           \\ \midrule
\textbf{Xception}          & 90.00\%                & 95.00\%                & 90.00\%          & 89.00\%           & {\ul 98.00\%}               & 93.00\%           \\ \midrule
\textbf{InceptionV3}       & 85.00\%                & 85.00\%                & 96.00\%          & 85.00\%           & \textbf{99.00\%}            & 82.00\%           \\ \midrule
\textbf{InceptionResNetV2} & 88.00\%                & 95.00\%                & 95.00\%          & 90.00\%           & 95.00\%                     & 93.00\%           \\ \midrule
\textbf{ResNet50}          & 95.00\%                & 89.00\%                & 94.00\%          & 96.00\%           & 95.00\%                     & 94.00\%           \\ \midrule
\textbf{ResNet152}         & 89.00\%                & 91.00\%                & 95.00\%          & 90.00\%           & 93.00\%                     & 89.00\%           \\ \midrule
\textbf{VGG16}             & 94.00\%                & 93.00\%                & 94.00\%          & 89.00\%           & 91.00\%                     & 89.00\%           \\ \midrule
\textbf{VGG19}             & 94.00\%                & 93.00\%                & 94.00\%          & 89.00\%           & 91.00\%                     & 89.00\%           \\ \midrule
\textbf{NASNetLarge}       & 88.00\%                & 91.00\%                & 94.00\%          & 90.00\%           & 95.00\%                     & 90.00\%           \\ \midrule
\textbf{NASNetMobile}      & 89.00\%                & 87.00\%                & 95.00\%          & 88.00\%           & 93.00\%                     & 85.00\%           \\ \midrule
\textbf{ResNet50V2}        & 91.00\%                & 89.00\%                & 94.00\%          & 88.00\%           & 96.00\%                     & 91.00\%           \\ \midrule
\textbf{ResNet101V2}       & 87.00\%                & 88.00\%                & 94.00\%          & 85.00\%           & 96.00\%                     & 77.00\%           \\ \midrule
\textbf{ResNet152V2}       & 90.00\%                & 94.00\%                & 96.00\%          & 90.00\%           & 96.00\%                     & 91.00\%           \\ \bottomrule
\end{tabular}%
}
\end{table}

\begin{table}[ht!]
\centering
\caption{Comparison of classification f1-score metric of different machine learning models. The bold value indicates the best result; underlined value represents the second-best result of the respective category.}
\label{tab:fscore}
\resizebox{\textwidth}{!}{%
\begin{tabular}{@{}lllllll@{}}
\toprule
\textbf{}                  & \textbf{Decision Tree} & \textbf{Random Forest} & \textbf{XGBoost} & \textbf{AdaBoost} & \textbf{Bagging Classifier} & \textbf{LightGBM} \\ \midrule
\textbf{MobileNet}         & 89.00\%                & 88.00\%                & 93.00\%          & 84.00\%           & \textbf{99.00\%}            & 91.00\%           \\ \midrule
\textbf{DesnseNet121}      & 96.00\%                & 96.00\%                & {\ul 98.00\%}    & 95.00\%           & 96.00\%                     & 95.00\%           \\ \midrule
\textbf{DenseNet201}       & 94.00\%                & 94.00\%                & 95.00\%          & 94.00\%           & {\ul 98.00\%}               & 94.00\%           \\ \midrule
\textbf{Xception}          & 90.00\%                & 95.00\%                & 90.00\%          & 89.00\%           & {\ul 98.00\%}               & 93.00\%           \\ \midrule
\textbf{InceptionV3}       & 85.00\%                & 85.00\%                & 96.00\%          & 85.00\%           & \textbf{99.00\%}            & 82.00\%           \\ \midrule
\textbf{InceptionResNetV2} & 88.00\%                & 95.00\%                & 95.00\%          & 90.00\%           & 95.00\%                     & 93.00\%           \\ \midrule
\textbf{ResNet50}          & 95.00\%                & 89.00\%                & 94.00\%          & 96.00\%           & 95.00\%                     & 94.00\%           \\ \midrule
\textbf{ResNet152}         & 89.00\%                & 91.00\%                & 95.00\%          & 90.00\%           & 93.00\%                     & 89.00\%           \\ \midrule
\textbf{VGG16}             & 94.00\%                & 93.00\%                & 94.00\%          & 89.00\%           & 91.00\%                     & 89.00\%           \\ \midrule
\textbf{VGG19}             & 94.00\%                & 93.00\%                & 94.00\%          & 89.00\%           & 91.00\%                     & 89.00\%           \\ \midrule
\textbf{NASNetLarge}       & 88.00\%                & 91.00\%                & 94.00\%          & 90.00\%           & 95.00\%                     & 90.00\%           \\ \midrule
\textbf{NASNetMobile}      & 89.00\%                & 87.00\%                & 95.00\%          & 88.00\%           & 93.00\%                     & 85.00\%           \\ \midrule
\textbf{ResNet50V2}        & 91.00\%                & 89.00\%                & 94.00\%          & 88.00\%           & 96.00\%                     & 91.00\%           \\ \midrule
\textbf{ResNet101V2}       & 87.00\%                & 88.00\%                & 94.00\%          & 85.00\%           & 96.00\%                     & 77.00\%           \\ \midrule
\textbf{ResNet152V2}       & 90.00\%                & 94.00\%                & 96.00\%          & 90.00\%           & 96.00\%                     & 91.00\%           \\ \bottomrule
\end{tabular}%
}
\end{table}
\begin{table}[ht!]
\centering
\caption{The time for feature extraction of deep CNN models and training on ML algorithms using Intel(R) Core (TM) i7-8700K 3.7 GHz processors with 32 GB RAM, Nvidia GeForce GTX 1080 Ti GPU with 11GB RAM.}
\label{tab:extractiontime}
\resizebox{\textwidth}{!}{%
\begin{tabular}{@{}llllllll@{}}
\toprule
                           & \textbf{Extraction Time (s)} & \textbf{DT (s)} & \textbf{RF (s)} & \textbf{XGBoost (s)} & \textbf{AdaBoost (s)} & \textbf{Bagging Classifier (s)} & \textbf{LightGBM (s)} \\ \midrule
\textbf{MobileNet}         & 8.803                        & 0.022           & 0.008           & 0.438                & 0.023                 & 33.535                          & 1.097                 \\ \midrule
\textbf{DesnseNet121}      & 9.306                        & 0.017           & 0.009           & 0.362                & 0.021                 & 30.748                          & 0.897                 \\ \midrule
\textbf{DenseNet201}       & 38.227                       & 0.035           & 0.009           & 0.684                & 0.034                 & 33.446                          & 1.573                 \\ \midrule
\textbf{Xception}          & 10.819                       & 0.042           & 0.009           & 0.787                & 0.044                 & 35.144                          & 1.612                 \\ \midrule
\textbf{InceptionV3}       & 11.825                       & 0.045           & 0.009           & 0.86                 & 0.048                 & 37.54                           & 1.98                  \\ \midrule
\textbf{InceptionResNetV2} & 14.151                       & 0.035           & 0.009           & 0.575                & 0.035                 & 33.562                          & 1.169                 \\ \midrule
\textbf{ResNet50}          & 10.206                       & 0.034           & 0.009           & 0.694                & 0.04                  & 33.232                          & 0.96                  \\ \midrule
\textbf{ResNet152}         & 15.769                       & 0.031           & 0.01            & 0.653                & 0.031                 & 32.347                          & 1.114                 \\ \midrule
\textbf{VGG16}             & 14.746                       & 0.009           & 0.008           & 0.2                  & 0.012                 & 29.51                           & 0.498                 \\ \midrule
\textbf{VGG19}             & 14.359                       & 0.01            & 0.008           & 0.2                  & 0.013                 & 29.336                          & 0.494                 \\ \midrule
\textbf{NASNetLarge}       & 13.131                       & 0.066           & 0.01            & 1.409                & 0.067                 & 38.337                          & 2.542                 \\ \midrule
\textbf{NASNetMobile}      & 7.786                        & 0.024           & 0.009           & 0.429                & 0.024                 & 32.782                          & 0.93                  \\ \midrule
\textbf{ResNet50V2}        & 10.204                       & 0.044           & 0.009           & 0.691                & 0.045                 & 34.369                          & 1.798                 \\ \midrule
\textbf{ResNet101V2}       & 12.435                       & 0.047           & 0.009           & 0.776                & 0.048                 & 0.9634                          & 1.577                 \\ \midrule
\textbf{ResNet152V2}       & 16.67                        & 0.031           & 0.009           & 0.73                 & 0.032                 & 34.56                           & 1.514                 \\ \bottomrule
\end{tabular}%
}
\end{table}

\begin{figure}[ht!]
\centering\includegraphics[width=0.6\linewidth]{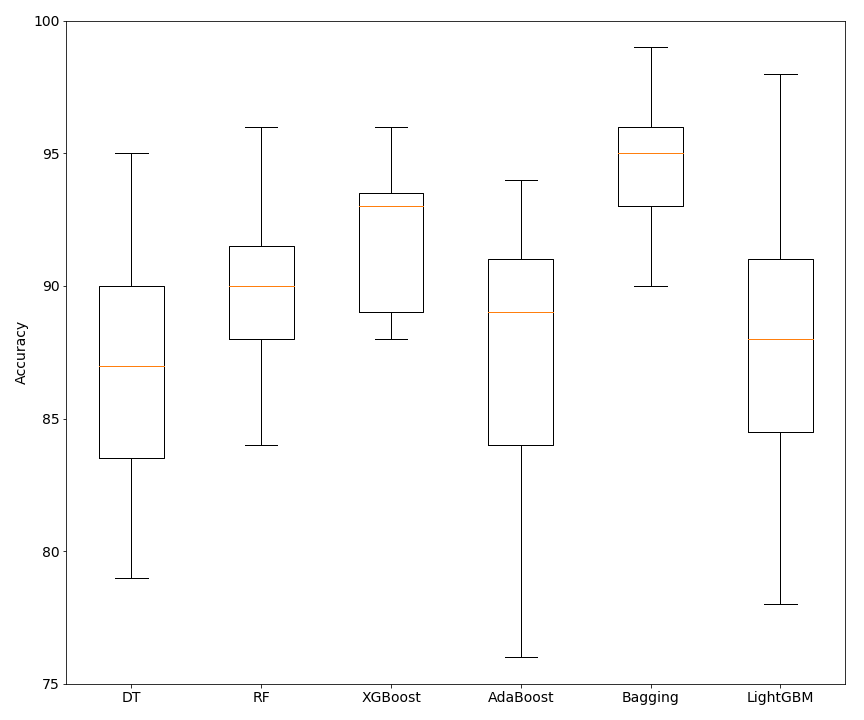}
\caption{Performance of different ML classifiers on the COVID-19 pneumonia classification.}
\label{fig:boxMLclassifiers}
\end{figure}

\begin{figure}[ht!]
\centering\includegraphics[width=0.8\linewidth]{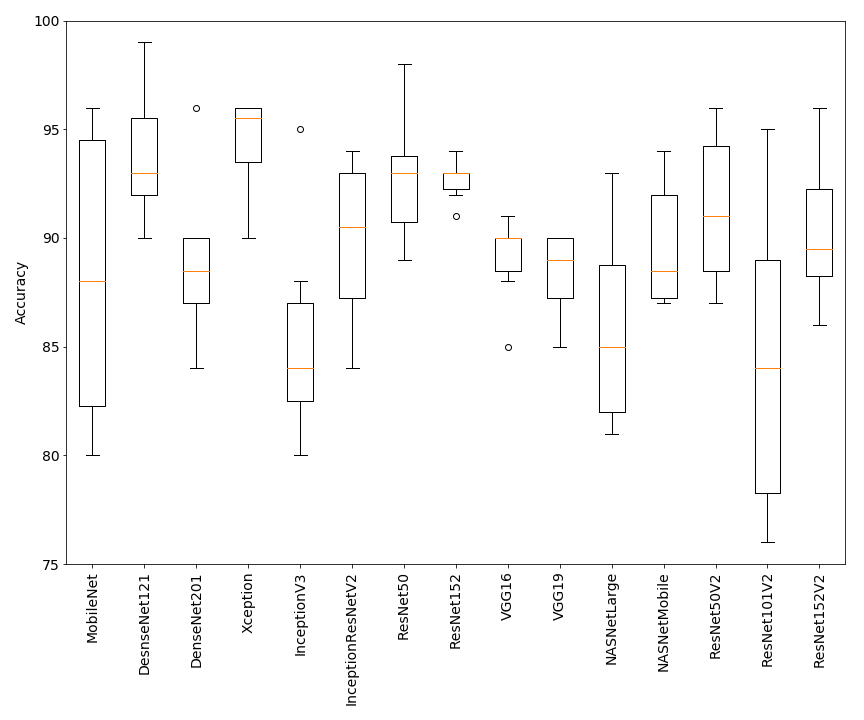}
\caption{Performance of the deep CNNs feature extractors and Bagging classifier on the COVID-19 pneumonia classification.}
\label{fig:boxCNNFeatures}
\end{figure}
\begin{figure}[ht!]
\centering\includegraphics[width=0.99\linewidth]{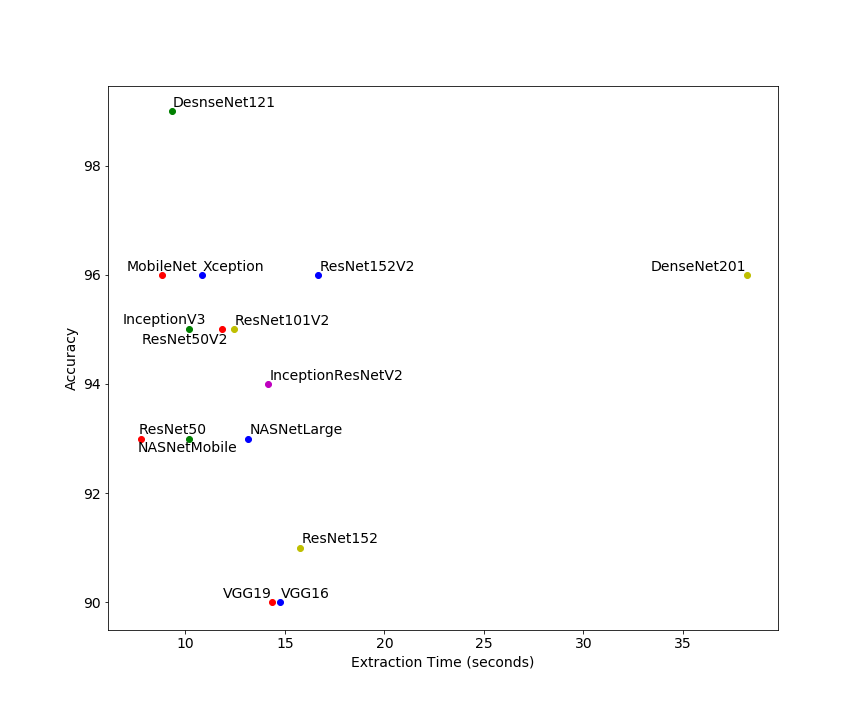}
\caption{Feature extraction time and accuracy of different deep CNN feature extractors on COVID-19 classification.}
\label{fig:extractiontime}
\end{figure}

\begin{figure}[ht!]
	\centering
	\begin{subfigure}{0.24\textwidth}
		\includegraphics[width=\linewidth]{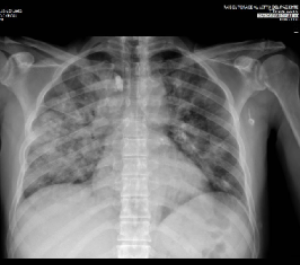}
% 		\caption{}
% 		\label{fig:original}
	\end{subfigure}\hfil
	\begin{subfigure}{0.24\textwidth}
		\includegraphics[width=\linewidth]{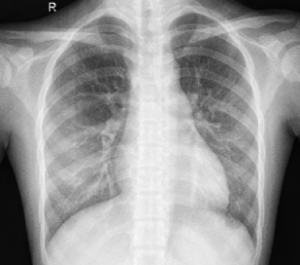}
% 		\caption{}
% 		\label{fig:VF}
	\end{subfigure}\hfil
	\begin{subfigure}{0.24\textwidth}
		\includegraphics[width=\linewidth]{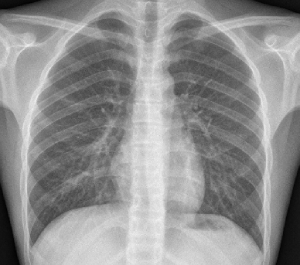}
% 		\caption{}
% 		\label{fig:HF}
	\end{subfigure}\hfil	
	\begin{subfigure}{0.24\textwidth}
		\includegraphics[width=\linewidth]{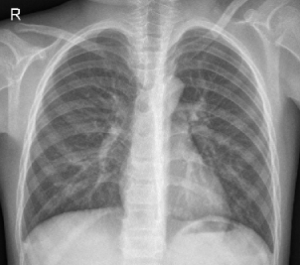}
% 		\caption{}
% 		\label{fig:RF}
	\end{subfigure}\hfil

	\caption{Examples of miss-classified cases of COVID-19 dataset.}
	\label{fig: MissClassified}
\end{figure}

Table~\ref{tab:accuracy} and Figure ~\ref{fig:boxMLclassifiers} summarize the accuracy performance of six machine learning algorithms, namely, DT, RF, XGBoost, AdaBoost, Bagging classifier and LightGBM on the feature extracted by deep CNNs. Each entry in Table~\ref{tab:accuracy}, is in the format ($\mu$ $\pm$ $\sigma$) where $\mu$ is the average classification accuracy and $\sigma$ is standard deviation. Analyzing Table~\ref{tab:accuracy} the topmost result was obtained by Bagging classifier  with a maximum of 99.00\% $\pm$ 0.09 accuracy on features extracted by DesnseNet121 architecture (with feature extraction time of 9.306 seconds and training time of 30.748 seconds in Table~\ref{tab:extractiontime}), which is the highest result reported in the literature for COVID-19 classification of this dataset. It is also inferred from Table~\ref{tab:accuracy} that the second-best result obtained by ResNet50 feature extractor and LightGBM classifier (with feature extraction time of 0.960 seconds and training time of 10.206 seconds in Table~\ref{tab:extractiontime}) with an overall accuracy of 98.00 $\pm$ 0.09. Comparing the first and second winners among all combinations, the classification accuracy of DenseNet121 with Bagging is slightly better (1\%) than ResNet50 with LightGBM, while the training time of the second winner is tempting, almost 30 times better than the first winner in terms of accuracy. Although Bagging is a slow learner, it has the lowest standard deviation and hence is more stable than other learners.

The results also demonstrate that the detection rate is worst on the features extracted by ResNet101V2 trained by the AdaBoost classifier with 76.00 $\pm$ 0.32 accuracy. Figure~\ref{fig:boxMLclassifiers} and Figure~\ref{fig:boxCNNFeatures} demonstrate box-plot distributions of deep CNNs feature extractors and classification accuracy from the 10-fold cross-validation. Circles in Figure~\ref{fig:boxMLclassifiers} represent outliers. In Tables~\ref{tab:precision},~\ref{tab:recall} and~\ref{tab:fscore}, the obtained precision, recall, and F1-score of the features extracted by deep CNN architectures trained by different learners are presented respectively. As given in these tables, the highest precision, recall, and F1-score rates are achieved by MobileNet and InceptionV3 feature vector trained on Bagging tree classifier with value of 99.00\% precision, recall, and F-score. The XGBoost and Bagging classifiers also yielded the second-best results with values of (98.00, 98.00, 98,00)\% precision, recall, and F-score rates with features extracted by DesnseNet121, DenseNet201 and Xception architectures. Similar conclusions can be drawn for other models. The experimental results indicate that the performance of the deep CNNs using DenseNet121, DenseNet201, MobileNet, Xception and InceptionV3 models trained by Bagging tree and XGBoost classifiers yield satisfactory results and outperforms other state-of-the-art CNNs and learners in COVID-19 classification. Based on the obtained results, we believe that by discarding the irrelevant features using sparse descriptors of low dimensionality features extracted by deep CNN models instead of training a deep CNNs model can be considered as a successful improvement of the performance of a machine learning algorithms. Our obtained results agree with the top-performing ML classifiers of Bagging and LightGBM. The best pre-trained visual feature extractor so far was DesnseNet121, MobileNet and InceptionV3 rather than counterpart architectures for COVID-19 image classification. 

Although the approach presented here shows satisfying performance, it also has limitations classifying more challenging instances with vague, low contrast boundaries, and the presence of artifacts. Some examples of these cases are illustrated in Figure~\ref{fig:extractiontime}. 
Finally, comparison of the feature extraction time using deep CNN models and training with ML algorithms are shown in Table~\ref{tab:extractiontime} and Figure~\ref{fig:extractiontime}. The extraction time of the DenseNet201 architectures on the total of 274 images was computed with 38.227 seconds (about 0.13 second per image) was the longest visual feature extractor and NASNetMobile was the fastest visual feature extractor by 7.786 seconds (about 0.028 second per image). DesnseNet121 architecture as the best model took 9.306 seconds (about 0.03 second per image) for feature extraction phase and 30.748 seconds (about 0.11 second per image) for the training phase on Bagging tree classifier. ResNet50 architecture as the second-best visual feature extractor took 10.206 seconds (about 0 0.03 second per image) for feature extraction phase and the training time of 0.960 seconds (about 0.003 second per image) on LightGBM classifier. In conclusion, the extraction and training time of the proposed approach is considerate significantly low in comparison with training a deep CNN model from scratch which implies faster computation time and lower resource consumption.

\begin{figure}[ht!]
	\centering
	\begin{subfigure}{0.4\textwidth}
		\includegraphics[width=\linewidth]{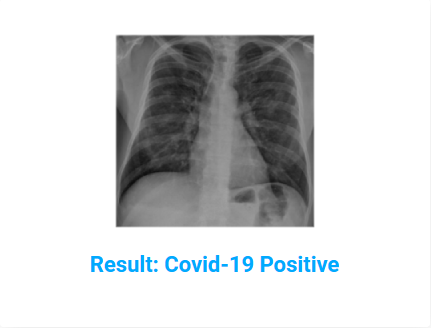}
% 		\caption{}
		\label{fig:original}
	\end{subfigure}\hfil
	\begin{subfigure}{0.4\textwidth}
		\includegraphics[width=\linewidth]{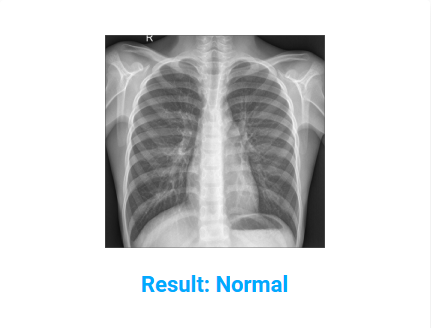}
% 		\caption{}
		\label{fig:VF}
	\end{subfigure}\hfil

	\caption{Web-based application for automatic detection of COVID-19 pneumonia.}
	\label{fig: webapp}
\end{figure}

After training a model, the pre-trained weights and models can be used as predictive engine for CAD systems to allow an automatic classification of new data.  A web-based application was implemented using standard web development tools and techniques such as Python, JavaScript, HTML, and Flask web framework. Figure~\ref{fig: webapp} shows the output of our web-based application for COVID-19 pneumonia detection. This web application could help doctors benefit from our proposed method by providing an online tool that only requires uploading an X-ray or CT image.  The application then provides the physician with a simple COVID-19 Positive, or COVID-19 Negative observation. It should be noted that this application has yet to be clinically validated, is not yet approved for diagnostic use and would simply serve as a diagnostic aid for the medical imaging specialist.

The proposed method is generic as it does not need handcrafted features and can be easily adapted, requiring minimal pre-processing. The provided dataset is collected across multiple sources with different shape, textures and morphological characteristics. The transfer learning strategy has successfully transferred knowledge from the source to the target domain despite the limited dataset size of the provided dataset. During the proposed approach, we observed that no overfitting occurs to impact the classification accuracy adversely.
However, our study has some limitations. The training data samples are limited. Extending the dataset size by additional data sources can provide a better understanding on the proposed approach. Also, employing pre-trained networks as feature extractors requires to rescale the input images to a certain dimension which may discard valuable information. Although the proposed methodology achieved satisfying performance with an accuracy of 99.00\%, the diagnostic performance of the deep learning visual feature extractor and machine learning classifier should be evaluated on real clinical study trials.

\section{Conclusion}
\label{S:Conclusion}
The ongoing pandemic of COVID-19 has been declared a global health emergency due to the relatively high infection rate of the disease. As of the time of this writing, there is no clinically approved therapeutic drug or vaccine available to treat COVID-19. Early detection of COVID-19 is important to interrupt the human-to-human transmission of COVID-19 and patient care. Currently, the isolation and quarantine of the suspicious patients is the most effective way to prevent the spread of COVID-19. Diagnostic modalities such as chest X-ray and CT are playing an important role in monitoring the progression and severity of the disease in COVID-19 positive patients. This paper presents a feature extractor-based deep learning and machine learning classifier approach for computer-aided diagnosis of COVID-19 pneumonia. Several ML algorithms were trained on the features extracted by well-established CNNs architectures to find the best combination of features and learners. Considering the high visual complexity of image data, proper deep feature extraction is considered as a critical step in developing deep CNN models. The experimental results on available chest X-ray and CT dataset demonstrate that the features extracted by DesnseNet121 architecture and trained by a Bagging tree classifier generates very accurate prediction of 99.00\% in terms of classification accuracy.

%% References
%%
%% Following citation commands can be used in the body text:
%% Usage of \cite is as follows:
%%   \cite{key}          ==>>  [#]
%%   \cite[chap. 2]{key} ==>>  [#, chap. 2]
%%   \citet{key}         ==>>  Author [#]

%% References with bibTeX database:

% \bibliographystyle{model1-num-names}

%% New version of the num-names style
\bibliographystyle{elsarticle-num-names}
\bibliography{reference.bib}

\begin{thebibliography}{46}
\expandafter\ifx\csname natexlab\endcsname\relax\def\natexlab#1{#1}\fi
\providecommand{\url}[1]{\texttt{#1}}
\providecommand{\href}[2]{#2}
\providecommand{\path}[1]{#1}
\providecommand{\DOIprefix}{doi:}
\providecommand{\ArXivprefix}{arXiv:}
\providecommand{\URLprefix}{URL: }
\providecommand{\Pubmedprefix}{pmid:}
\providecommand{\doi}[1]{\href{http://dx.doi.org/#1}{\path{#1}}}
\providecommand{\Pubmed}[1]{\href{pmid:#1}{\path{#1}}}
\providecommand{\bibinfo}[2]{#2}
\ifx\xfnm\relax \def\xfnm[#1]{\unskip,\space#1}\fi
%Type = Article
\bibitem[{Shereen et~al.(2020)Shereen, Khan, Kazmi, Bashir, and
  Siddique}]{SHEREEN202091}
\bibinfo{author}{M.~A. Shereen}, \bibinfo{author}{S.~Khan},
  \bibinfo{author}{A.~Kazmi}, \bibinfo{author}{N.~Bashir},
  \bibinfo{author}{R.~Siddique},
\newblock \bibinfo{title}{Covid-19 infection: Origin, transmission, and
  characteristics of human coronaviruses},
\newblock \bibinfo{journal}{Journal of Advanced Research} \bibinfo{volume}{24}
  (\bibinfo{year}{2020}) \bibinfo{pages}{91 -- 98}. \URLprefix
  \url{http://www.sciencedirect.com/science/article/pii/S2090123220300540}.
  \DOIprefix\doi{https://doi.org/10.1016/j.jare.2020.03.005}.
%Type = Article
\bibitem[{Lippi et~al.(2020)Lippi, Plebani, and Henry}]{LIPPI2020145}
\bibinfo{author}{G.~Lippi}, \bibinfo{author}{M.~Plebani},
  \bibinfo{author}{B.~M. Henry},
\newblock \bibinfo{title}{Thrombocytopenia is associated with severe
  coronavirus disease 2019 (covid-19) infections: A meta-analysis},
\newblock \bibinfo{journal}{Clinica Chimica Acta} \bibinfo{volume}{506}
  (\bibinfo{year}{2020}) \bibinfo{pages}{145 -- 148}. \URLprefix
  \url{http://www.sciencedirect.com/science/article/pii/S0009898120301248}.
  \DOIprefix\doi{https://doi.org/10.1016/j.cca.2020.03.022}.
%Type = Article
\bibitem[{Zhang et~al.(2020)Zhang, Wu, and Zhang}]{ZHANG20201346}
\bibinfo{author}{T.~Zhang}, \bibinfo{author}{Q.~Wu},
  \bibinfo{author}{Z.~Zhang},
\newblock \bibinfo{title}{Probable pangolin origin of sars-cov-2 associated
  with the covid-19 outbreak},
\newblock \bibinfo{journal}{Current Biology} \bibinfo{volume}{30}
  (\bibinfo{year}{2020}) \bibinfo{pages}{1346 -- 1351.e2}. \URLprefix
  \url{http://www.sciencedirect.com/science/article/pii/S0960982220303602}.
  \DOIprefix\doi{https://doi.org/10.1016/j.cub.2020.03.022}.
%Type = Article
\bibitem[{Guo et~al.(2020)Guo, Zhou, Liu, and Tan}]{GUO2020}
\bibinfo{author}{H.~Guo}, \bibinfo{author}{Y.~Zhou}, \bibinfo{author}{X.~Liu},
  \bibinfo{author}{J.~Tan},
\newblock \bibinfo{title}{The impact of the covid-19 epidemic on the
  utilization of emergency dental services},
\newblock \bibinfo{journal}{Journal of Dental Sciences}
  (\bibinfo{year}{2020}). \URLprefix
  \url{http://www.sciencedirect.com/science/article/pii/S1991790220300209}.
  \DOIprefix\doi{https://doi.org/10.1016/j.jds.2020.02.002}.
%Type = Article
\bibitem[{Chavez et~al.(2020)Chavez, Long, Koyfman, and Liang}]{CHAVEZ2020}
\bibinfo{author}{S.~Chavez}, \bibinfo{author}{B.~Long},
  \bibinfo{author}{A.~Koyfman}, \bibinfo{author}{S.~Y. Liang},
\newblock \bibinfo{title}{Coronavirus disease (covid-19): A primer for
  emergency physicians},
\newblock \bibinfo{journal}{The American Journal of Emergency Medicine}
  (\bibinfo{year}{2020}). \URLprefix
  \url{http://www.sciencedirect.com/science/article/pii/S0735675720301789}.
  \DOIprefix\doi{https://doi.org/10.1016/j.ajem.2020.03.036}.
%Type = Article
\bibitem[{Rothan and Byrareddy(2020)}]{ROTHAN2020102433}
\bibinfo{author}{H.~A. Rothan}, \bibinfo{author}{S.~N. Byrareddy},
\newblock \bibinfo{title}{The epidemiology and pathogenesis of coronavirus
  disease (covid-19) outbreak},
\newblock \bibinfo{journal}{Journal of Autoimmunity} \bibinfo{volume}{109}
  (\bibinfo{year}{2020}) \bibinfo{pages}{102433}. \URLprefix
  \url{http://www.sciencedirect.com/science/article/pii/S0896841120300469}.
  \DOIprefix\doi{https://doi.org/10.1016/j.jaut.2020.102433}.
%Type = Article
\bibitem[{Liu et~al.(2020)Liu, Liu, Li, Zhang, Wang, and Lan}]{LIU2020}
\bibinfo{author}{H.~Liu}, \bibinfo{author}{F.~Liu}, \bibinfo{author}{J.~Li},
  \bibinfo{author}{T.~Zhang}, \bibinfo{author}{D.~Wang},
  \bibinfo{author}{W.~Lan},
\newblock \bibinfo{title}{Clinical and ct imaging features of the covid-19
  pneumonia: Focus on pregnant women and children},
\newblock \bibinfo{journal}{Journal of Infection}  (\bibinfo{year}{2020}).
  \URLprefix
  \url{http://www.sciencedirect.com/science/article/pii/S0163445320301183}.
  \DOIprefix\doi{https://doi.org/10.1016/j.jinf.2020.03.007}.
%Type = Misc
\bibitem[{{WHO}(2020)}]{CoronavirusStat}
\bibinfo{author}{{WHO}}, \bibinfo{title}{{Coronavirus disease (COVID-19)
  Pandemic}}, \bibinfo{year}{2020}. \URLprefix
  \url{https://www.who.int/emergencies/diseases/novel-coronavirus-2019}.
%Type = Article
\bibitem[{Shim et~al.(2020)Shim, Tariq, Choi, Lee, and Chowell}]{SHIM2020339}
\bibinfo{author}{E.~Shim}, \bibinfo{author}{A.~Tariq},
  \bibinfo{author}{W.~Choi}, \bibinfo{author}{Y.~Lee},
  \bibinfo{author}{G.~Chowell},
\newblock \bibinfo{title}{Transmission potential and severity of covid-19 in
  south korea},
\newblock \bibinfo{journal}{International Journal of Infectious Diseases}
  \bibinfo{volume}{93} (\bibinfo{year}{2020}) \bibinfo{pages}{339 -- 344}.
  \URLprefix
  \url{http://www.sciencedirect.com/science/article/pii/S1201971220301508}.
  \DOIprefix\doi{https://doi.org/10.1016/j.ijid.2020.03.031}.
%Type = Misc
\bibitem[{{CDC}(2020{\natexlab{a}})}]{cdcWebLink1}
\bibinfo{author}{{CDC}}, \bibinfo{title}{{Coronavirus Infections}},
  \bibinfo{year}{2020}{\natexlab{a}}. \URLprefix
  \url{https://phil.cdc.gov/Details.aspx?pid=23313}.
%Type = Misc
\bibitem[{{CDC}(2020{\natexlab{b}})}]{cdcWebLink2}
\bibinfo{author}{{CDC}}, \bibinfo{title}{{Coronavirus Infections - Transmission
  electron microscopic image}}, \bibinfo{year}{2020}{\natexlab{b}}. \URLprefix
  \url{https://phil.cdc.gov/Details.aspx?pid=23354}.
%Type = Article
\bibitem[{Wang et~al.(2020)Wang, Dong, Hu, Li, Ren, Zhang, Shi, and
  Zhou}]{wang2020temporal}
\bibinfo{author}{Y.~Wang}, \bibinfo{author}{C.~Dong}, \bibinfo{author}{Y.~Hu},
  \bibinfo{author}{C.~Li}, \bibinfo{author}{Q.~Ren},
  \bibinfo{author}{X.~Zhang}, \bibinfo{author}{H.~Shi},
  \bibinfo{author}{M.~Zhou},
\newblock \bibinfo{title}{Temporal changes of ct findings in 90 patients with
  covid-19 pneumonia: a longitudinal study},
\newblock \bibinfo{journal}{Radiology}  (\bibinfo{year}{2020})
  \bibinfo{pages}{200843}.
%Type = Article
\bibitem[{Hemdan et~al.(2020)Hemdan, Shouman, and Karar}]{hemdan2020covidx}
\bibinfo{author}{E.~E.-D. Hemdan}, \bibinfo{author}{M.~A. Shouman},
  \bibinfo{author}{M.~E. Karar},
\newblock \bibinfo{title}{Covidx-net: A framework of deep learning classifiers
  to diagnose covid-19 in x-ray images},
\newblock \bibinfo{journal}{arXiv preprint arXiv:2003.11055}
  (\bibinfo{year}{2020}).
%Type = Article
\bibitem[{Cohen et~al.(2020)Cohen, Morrison, and Dao}]{cohen2020covid}
\bibinfo{author}{J.~P. Cohen}, \bibinfo{author}{P.~Morrison},
  \bibinfo{author}{L.~Dao},
\newblock \bibinfo{title}{Covid-19 image data collection},
\newblock \bibinfo{journal}{arXiv 2003.11597}  (\bibinfo{year}{2020}).
  \URLprefix \url{https://github.com/ieee8023/covid-chestxray-dataset}.
%Type = Misc
\bibitem[{{Adrian Rosebrock}(2020)}]{AdrianRosebrock}
\bibinfo{author}{{Adrian Rosebrock}}, \bibinfo{title}{{Detecting COVID-19 in
  X-ray images with Keras, TensorFlow, and Deep Learning}},
  \bibinfo{year}{2020}. \URLprefix
  \url{https://www.pyimagesearch.com/2020/03/16/detecting-covid-19-in-x-ray-images-with-keras-tensorflow-and-deep-learning/}.
%Type = Article
\bibitem[{Barstugan et~al.(2020)Barstugan, Ozkaya, and
  Ozturk}]{barstugan2020coronavirus}
\bibinfo{author}{M.~Barstugan}, \bibinfo{author}{U.~Ozkaya},
  \bibinfo{author}{S.~Ozturk},
\newblock \bibinfo{title}{Coronavirus (covid-19) classification using ct images
  by machine learning methods},
\newblock \bibinfo{journal}{arXiv preprint arXiv:2003.09424}
  (\bibinfo{year}{2020}).
%Type = Book
\bibitem[{Cristianini et~al.(2000)Cristianini, Shawe-Taylor
  et~al.}]{cristianini2000introduction}
\bibinfo{author}{N.~Cristianini}, \bibinfo{author}{J.~Shawe-Taylor}, et~al.,
  \bibinfo{title}{An introduction to support vector machines and other
  kernel-based learning methods}, \bibinfo{publisher}{Cambridge university
  press}, \bibinfo{year}{2000}.
%Type = Article
\bibitem[{Wang and Wong(2020)}]{wang2020covid}
\bibinfo{author}{L.~Wang}, \bibinfo{author}{A.~Wong},
\newblock \bibinfo{title}{Covid-net: A tailored deep convolutional neural
  network design for detection of covid-19 cases from chest radiography
  images},
\newblock \bibinfo{journal}{arXiv preprint arXiv:2003.09871}
  (\bibinfo{year}{2020}).
%Type = Misc
\bibitem[{{Kaggle}(2020)}]{kagglechestxray}
\bibinfo{author}{{Kaggle}}, \bibinfo{title}{{Kaggle's Chest X-Ray Images
  (Pneumonia) dataset}}, \bibinfo{year}{2020}. \URLprefix
  \url{https://www.kaggle.com/paultimothymooney/chest-xray-pneumonia}.
%Type = Article
\bibitem[{Maghdid et~al.(2020)Maghdid, Asaad, Ghafoor, Sadiq, and
  Khan}]{maghdid2020diagnosing}
\bibinfo{author}{H.~S. Maghdid}, \bibinfo{author}{A.~T. Asaad},
  \bibinfo{author}{K.~Z. Ghafoor}, \bibinfo{author}{A.~S. Sadiq},
  \bibinfo{author}{M.~K. Khan},
\newblock \bibinfo{title}{Diagnosing covid-19 pneumonia from x-ray and ct
  images using deep learning and transfer learning algorithms},
\newblock \bibinfo{journal}{arXiv preprint arXiv:2004.00038}
  (\bibinfo{year}{2020}).
%Type = Inproceedings
\bibitem[{Krizhevsky et~al.(2012)Krizhevsky, Sutskever, and
  Hinton}]{krizhevsky2012imagenet}
\bibinfo{author}{A.~Krizhevsky}, \bibinfo{author}{I.~Sutskever},
  \bibinfo{author}{G.~E. Hinton},
\newblock \bibinfo{title}{Imagenet classification with deep convolutional
  neural networks},
\newblock in: \bibinfo{booktitle}{Advances in neural information processing
  systems}, \bibinfo{year}{2012}, pp. \bibinfo{pages}{1097--1105}.
%Type = Article
\bibitem[{Ghoshal and Tucker(2020)}]{ghoshal2020estimating}
\bibinfo{author}{B.~Ghoshal}, \bibinfo{author}{A.~Tucker},
\newblock \bibinfo{title}{Estimating uncertainty and interpretability in deep
  learning for coronavirus (covid-19) detection},
\newblock \bibinfo{journal}{arXiv preprint arXiv:2003.10769}
  (\bibinfo{year}{2020}).
%Type = Article
\bibitem[{Hall et~al.(2020)Hall, Paul, Goldgof, and Goldgof}]{hall2020finding}
\bibinfo{author}{L.~O. Hall}, \bibinfo{author}{R.~Paul}, \bibinfo{author}{D.~B.
  Goldgof}, \bibinfo{author}{G.~M. Goldgof},
\newblock \bibinfo{title}{Finding covid-19 from chest x-rays using deep
  learning on a small dataset},
\newblock \bibinfo{journal}{arXiv preprint arXiv:2004.02060}
  (\bibinfo{year}{2020}).
%Type = Article
\bibitem[{Farooq and Hafeez(2020)}]{farooq2020covid}
\bibinfo{author}{M.~Farooq}, \bibinfo{author}{A.~Hafeez},
\newblock \bibinfo{title}{Covid-resnet: A deep learning framework for screening
  of covid19 from radiographs},
\newblock \bibinfo{journal}{arXiv preprint arXiv:2003.14395}
  (\bibinfo{year}{2020}).
%Type = Inproceedings
\bibitem[{Boureau et~al.(2010)Boureau, Ponce, and
  LeCun}]{boureau2010theoretical}
\bibinfo{author}{Y.-L. Boureau}, \bibinfo{author}{J.~Ponce},
  \bibinfo{author}{Y.~LeCun},
\newblock \bibinfo{title}{A theoretical analysis of feature pooling in visual
  recognition},
\newblock in: \bibinfo{booktitle}{Proceedings of the 27th international
  conference on machine learning (ICML-10)}, \bibinfo{year}{2010}, pp.
  \bibinfo{pages}{111--118}.
%Type = Inproceedings
\bibitem[{Rakhlin et~al.(2018)Rakhlin, Shvets, Iglovikov, and
  Kalinin}]{rakhlin2018deep}
\bibinfo{author}{A.~Rakhlin}, \bibinfo{author}{A.~Shvets},
  \bibinfo{author}{V.~Iglovikov}, \bibinfo{author}{A.~A. Kalinin},
\newblock \bibinfo{title}{Deep convolutional neural networks for breast cancer
  histology image analysis},
\newblock in: \bibinfo{booktitle}{International Conference Image Analysis and
  Recognition}, \bibinfo{organization}{Springer}, \bibinfo{year}{2018}, pp.
  \bibinfo{pages}{737--744}.
%Type = Article
\bibitem[{Guo et~al.(2016)Guo, Liu, Oerlemans, Lao, Wu, and Lew}]{guo2016deep}
\bibinfo{author}{Y.~Guo}, \bibinfo{author}{Y.~Liu},
  \bibinfo{author}{A.~Oerlemans}, \bibinfo{author}{S.~Lao},
  \bibinfo{author}{S.~Wu}, \bibinfo{author}{M.~S. Lew},
\newblock \bibinfo{title}{Deep learning for visual understanding: A review},
\newblock \bibinfo{journal}{Neurocomputing} \bibinfo{volume}{187}
  (\bibinfo{year}{2016}) \bibinfo{pages}{27--48}.
%Type = Article
\bibitem[{Quinlan(1986)}]{quinlan1986induction}
\bibinfo{author}{J.~R. Quinlan},
\newblock \bibinfo{title}{Induction of decision trees},
\newblock \bibinfo{journal}{Machine learning} \bibinfo{volume}{1}
  (\bibinfo{year}{1986}) \bibinfo{pages}{81--106}.
%Type = Article
\bibitem[{Breiman(2001)}]{Breiman2001}
\bibinfo{author}{L.~Breiman},
\newblock \bibinfo{title}{Random forests},
\newblock \bibinfo{journal}{Machine learning} \bibinfo{volume}{45}
  (\bibinfo{year}{2001}) \bibinfo{pages}{5--32}.
%Type = Inproceedings
\bibitem[{Chen and Guestrin(2016)}]{Chen2016}
\bibinfo{author}{T.~Chen}, \bibinfo{author}{C.~Guestrin},
\newblock \bibinfo{title}{{XGBoost}},
\newblock in: \bibinfo{booktitle}{Proceedings of the 22nd ACM SIGKDD
  International Conference on Knowledge Discovery and Data Mining - KDD '16},
  \bibinfo{publisher}{ACM Press}, \bibinfo{address}{New York, New York, USA},
  \bibinfo{year}{2016}, pp. \bibinfo{pages}{785--794}. \URLprefix
  \url{http://dl.acm.org/citation.cfm?doid=2939672.2939785}.
  \DOIprefix\doi{10.1145/2939672.2939785}.
%Type = Inproceedings
\bibitem[{Freund and Schapire(1995)}]{freund1995desicion}
\bibinfo{author}{Y.~Freund}, \bibinfo{author}{R.~E. Schapire},
\newblock \bibinfo{title}{A desicion-theoretic generalization of on-line
  learning and an application to boosting},
\newblock in: \bibinfo{booktitle}{European conference on computational learning
  theory}, \bibinfo{organization}{Springer}, \bibinfo{year}{1995}, pp.
  \bibinfo{pages}{23--37}.
%Type = Article
\bibitem[{Breiman(1996)}]{breiman1996bagging}
\bibinfo{author}{L.~Breiman},
\newblock \bibinfo{title}{Bagging predictors},
\newblock \bibinfo{journal}{Machine learning} \bibinfo{volume}{24}
  (\bibinfo{year}{1996}) \bibinfo{pages}{123--140}.
%Type = Inproceedings
\bibitem[{Ke et~al.(2017)Ke, Meng, Finley, Wang, Chen, Ma, Ye, and
  Liu}]{ke2017lightgbm}
\bibinfo{author}{G.~Ke}, \bibinfo{author}{Q.~Meng},
  \bibinfo{author}{T.~Finley}, \bibinfo{author}{T.~Wang},
  \bibinfo{author}{W.~Chen}, \bibinfo{author}{W.~Ma}, \bibinfo{author}{Q.~Ye},
  \bibinfo{author}{T.-Y. Liu},
\newblock \bibinfo{title}{Lightgbm: A highly efficient gradient boosting
  decision tree},
\newblock in: \bibinfo{booktitle}{Advances in neural information processing
  systems}, \bibinfo{year}{2017}, pp. \bibinfo{pages}{3146--3154}.
%Type = Article
\bibitem[{Kassani et~al.(2019)Kassani, Kassani, Wesolowski, Schneider, and
  Deters}]{kassani2019breast}
\bibinfo{author}{S.~H. Kassani}, \bibinfo{author}{P.~H. Kassani},
  \bibinfo{author}{M.~J. Wesolowski}, \bibinfo{author}{K.~A. Schneider},
  \bibinfo{author}{R.~Deters},
\newblock \bibinfo{title}{Breast cancer diagnosis with transfer learning and
  global pooling},
\newblock \bibinfo{journal}{arXiv preprint arXiv:1909.11839}
  (\bibinfo{year}{2019}).
%Type = Article
\bibitem[{Khan et~al.(2019)Khan, Islam, Jan, Din, and
  Rodrigues}]{khan2019novel}
\bibinfo{author}{S.~Khan}, \bibinfo{author}{N.~Islam},
  \bibinfo{author}{Z.~Jan}, \bibinfo{author}{I.~U. Din},
  \bibinfo{author}{J.~J.~C. Rodrigues},
\newblock \bibinfo{title}{A novel deep learning based framework for the
  detection and classification of breast cancer using transfer learning},
\newblock \bibinfo{journal}{Pattern Recognition Letters} \bibinfo{volume}{125}
  (\bibinfo{year}{2019}) \bibinfo{pages}{1--6}.
%Type = Article
\bibitem[{Mehra et~al.(2018)}]{mehra2018breast}
\bibinfo{author}{R.~Mehra}, et~al.,
\newblock \bibinfo{title}{Breast cancer histology images classification:
  Training from scratch or transfer learning?},
\newblock \bibinfo{journal}{ICT Express} \bibinfo{volume}{4}
  (\bibinfo{year}{2018}) \bibinfo{pages}{247--254}.
%Type = Article
\bibitem[{Lu et~al.(2019)Lu, Lu, and Zhang}]{lu2019pathological}
\bibinfo{author}{S.~Lu}, \bibinfo{author}{Z.~Lu}, \bibinfo{author}{Y.-D.
  Zhang},
\newblock \bibinfo{title}{Pathological brain detection based on alexnet and
  transfer learning},
\newblock \bibinfo{journal}{Journal of computational science}
  \bibinfo{volume}{30} (\bibinfo{year}{2019}) \bibinfo{pages}{41--47}.
%Type = Inproceedings
\bibitem[{Kassani et~al.(2019)Kassani, Kassani, Wesolowski, Schneider, and
  Deters}]{biopsybinary}
\bibinfo{author}{S.~H. Kassani}, \bibinfo{author}{P.~H. Kassani},
  \bibinfo{author}{M.~J. Wesolowski}, \bibinfo{author}{K.~A. Schneider},
  \bibinfo{author}{R.~Deters},
\newblock \bibinfo{title}{Classification of histopathological biopsy images
  using ensemble of deep learning networks},
\newblock in: \bibinfo{booktitle}{Proceedings of the 29th Annual International
  Conference on Computer Science and Software Engineering}, CASCON ’19,
  \bibinfo{publisher}{IBM Corp.}, \bibinfo{address}{USA}, \bibinfo{year}{2019},
  p. \bibinfo{pages}{92–99}.
%Type = Article
\bibitem[{Liu et~al.(2020)Liu, Cao, Li, Xiao, Qiu, Yang, Zhao, and
  Cui}]{liu2020automatic}
\bibinfo{author}{Z.~Liu}, \bibinfo{author}{Y.~Cao}, \bibinfo{author}{Y.~Li},
  \bibinfo{author}{X.~Xiao}, \bibinfo{author}{Q.~Qiu},
  \bibinfo{author}{M.~Yang}, \bibinfo{author}{Y.~Zhao},
  \bibinfo{author}{L.~Cui},
\newblock \bibinfo{title}{Automatic diagnosis of fungal keratitis using data
  augmentation and image fusion with deep convolutional neural network},
\newblock \bibinfo{journal}{Computer Methods and Programs in Biomedicine}
  \bibinfo{volume}{187} (\bibinfo{year}{2020}) \bibinfo{pages}{105019}.
%Type = Article
\bibitem[{Liu et~al.(2019)Liu, Tian, and Xu}]{liu2019novel}
\bibinfo{author}{S.~Liu}, \bibinfo{author}{G.~Tian}, \bibinfo{author}{Y.~Xu},
\newblock \bibinfo{title}{A novel scene classification model combining resnet
  based transfer learning and data augmentation with a filter},
\newblock \bibinfo{journal}{Neurocomputing} \bibinfo{volume}{338}
  (\bibinfo{year}{2019}) \bibinfo{pages}{191--206}.
%Type = Article
\bibitem[{Sridar et~al.(2019)Sridar, Kumar, Quinton, Nanan, Kim, and
  Krishnakumar}]{sridar2019decision}
\bibinfo{author}{P.~Sridar}, \bibinfo{author}{A.~Kumar},
  \bibinfo{author}{A.~Quinton}, \bibinfo{author}{R.~Nanan},
  \bibinfo{author}{J.~Kim}, \bibinfo{author}{R.~Krishnakumar},
\newblock \bibinfo{title}{Decision fusion-based fetal ultrasound image plane
  classification using convolutional neural networks},
\newblock \bibinfo{journal}{Ultrasound in medicine \& biology}
  \bibinfo{volume}{45} (\bibinfo{year}{2019}) \bibinfo{pages}{1259--1273}.
%Type = Article
\bibitem[{Zhang et~al.(2019)Zhang, Zhong, Yang, Gao, Hu, Chen, and
  Yi}]{zhang2019automated}
\bibinfo{author}{W.~Zhang}, \bibinfo{author}{J.~Zhong},
  \bibinfo{author}{S.~Yang}, \bibinfo{author}{Z.~Gao}, \bibinfo{author}{J.~Hu},
  \bibinfo{author}{Y.~Chen}, \bibinfo{author}{Z.~Yi},
\newblock \bibinfo{title}{Automated identification and grading system of
  diabetic retinopathy using deep neural networks},
\newblock \bibinfo{journal}{Knowledge-Based Systems} \bibinfo{volume}{175}
  (\bibinfo{year}{2019}) \bibinfo{pages}{12--25}.
%Type = Article
\bibitem[{Dourado~Jr et~al.(2019)Dourado~Jr, da~Silva, da~N{\'o}brega, Barros,
  Reboucas~Filho, and de~Albuquerque}]{dourado2019deep}
\bibinfo{author}{C.~M. Dourado~Jr}, \bibinfo{author}{S.~P.~P. da~Silva},
  \bibinfo{author}{R.~V.~M. da~N{\'o}brega}, \bibinfo{author}{A.~C. d.~S.
  Barros}, \bibinfo{author}{P.~P. Reboucas~Filho}, \bibinfo{author}{V.~H.~C.
  de~Albuquerque},
\newblock \bibinfo{title}{Deep learning iot system for online stroke detection
  in skull computed tomography images},
\newblock \bibinfo{journal}{Computer Networks} \bibinfo{volume}{152}
  (\bibinfo{year}{2019}) \bibinfo{pages}{25--39}.
%Type = Article
\bibitem[{{\c{C}}inar and Y{\i}ld{\i}r{\i}m(2020)}]{ccinar2020detection}
\bibinfo{author}{A.~{\c{C}}inar}, \bibinfo{author}{M.~Y{\i}ld{\i}r{\i}m},
\newblock \bibinfo{title}{Detection of tumors on brain mri images using the
  hybrid convolutional neural network architecture},
\newblock \bibinfo{journal}{Medical Hypotheses}  (\bibinfo{year}{2020})
  \bibinfo{pages}{109684}.
%Type = Article
\bibitem[{Cogan et~al.(2019)Cogan, Cogan, and Tamil}]{cogan2019mapgi}
\bibinfo{author}{T.~Cogan}, \bibinfo{author}{M.~Cogan},
  \bibinfo{author}{L.~Tamil},
\newblock \bibinfo{title}{Mapgi: Accurate identification of anatomical
  landmarks and diseased tissue in gastrointestinal tract using deep learning},
\newblock \bibinfo{journal}{Computers in biology and medicine}
  \bibinfo{volume}{111} (\bibinfo{year}{2019}) \bibinfo{pages}{103351}.
%Type = Misc
\bibitem[{{Kaggle}(2020)}]{kaggleRSNA}
\bibinfo{author}{{Kaggle}}, \bibinfo{title}{{RSNA Pneumonia Detection
  Challenge}}, \bibinfo{year}{2020}. \URLprefix
  \url{https://www.kaggle.com/c/rsna-pneumonia-detection-challenge}.

\end{thebibliography}

%% Authors are advised to submit their bibtex database files. They are
%% requested to list a bibtex style file in the manuscript if they do
%% not want to use model1-num-names.bst.

%% References without bibTeX database:

% \begin{thebibliography}{00}

%% \bibitem must have the following form:
%%   \bibitem{key}...
%%

% \bibitem{}

% \end{thebibliography}

\end{document}